\documentclass[twocolumn,prl]{revtex4}

\ifx\pdfoutput\undefined
  \usepackage[dvips]{graphicx}
\else
  \usepackage[pdftex]{graphicx}
\fi

\usepackage{dcolumn}
\usepackage{amsmath}
\usepackage{latexsym}

\begin{document}

\title[Short Title]{Josephson charge-phase qubit with radio
frequency readout: \\ coupling and decoherence}

\author{A.~B.~Zorin}
 \email{alexander.zorin@ptb.de}

\affiliation{Physikalisch-Technische Bundesanstalt, Bundesallee
100, 38116 Braunschweig, Germany}%

\date{9 December, 2003}  

\begin{abstract}
The Josephson charge-phase qubit based on a superconducting
single charge transistor inserted in a low-inductance
superconducting loop is considered. The loop is inductively
coupled to a radio-frequency driven tank circuit enabling the
readout of the qubit states by measuring the effective Josephson
inductance of the transistor. The effect of qubit dephasing and
relaxation due to electric and magnetic control lines as well as
the measuring system is evaluated. Recommendations for the qubit
operation in the magic points with minimum decoherence are given.


\verb  PACS numbers: 74.50.+r, 85.25.Cp, 03.67.Lx

\end{abstract}

\maketitle

\section{Introduction}

The superconducting quantum bit (qubit) circuits comprising
mesoscopic Josephson tunnel junctions have recently demonstrated
remarkable quantum coherence properties and now are considered to
be promising elements for a scalable quantum computer
\cite{JClarke}. The readout of macroscopic quantum states of a
single qubit or a system of coupled qubits with minimum
decoherence being caused by the detector, remains, however, one
of the most important engineering issues in this field.

The Josephson qubits are commonly subdivided into flux, phase,
charge and charge-phase qubits. The design of charge and
charge-phase qubits is based on a Cooper pair box \cite{Bouchiat}
in which a small superconducting island with significant Coulomb
energy is charged through a small Josephson junction (charge
qubit) or a miniature double-junction SQUID (charge-phase). The
distinct quantum states of the box generated by signals applied to
a gate are associated with different observable charges on the
island. This makes it possible to read out the qubit state by
discriminating the island charge. Probing this charge can be done
either by single quasiparticle tunneling across a small auxiliary
tunnel junction attached to the island \cite{Nakamura} or by a
capacitively coupled electrometer \cite{CTH-rf-SET}. In the
charge-phase qubits, the quantum states of the box involve the
phase coordinate of the SQUID loop, so discriminating these states
can also be done by measuring the persistent current circulating
in the loop at an appropriate dc flux bias. Such a measurement was
performed in the experiment of the Saclay group \cite{Vion}. In
their setup, nicknamed "Quantronium", the circulating current
passed through a larger auxiliary (third) junction was read out
by measuring the switching current of this junction.

\begin{figure}[b]
\begin{center}
\leavevmode
\includegraphics[width=3.1in]{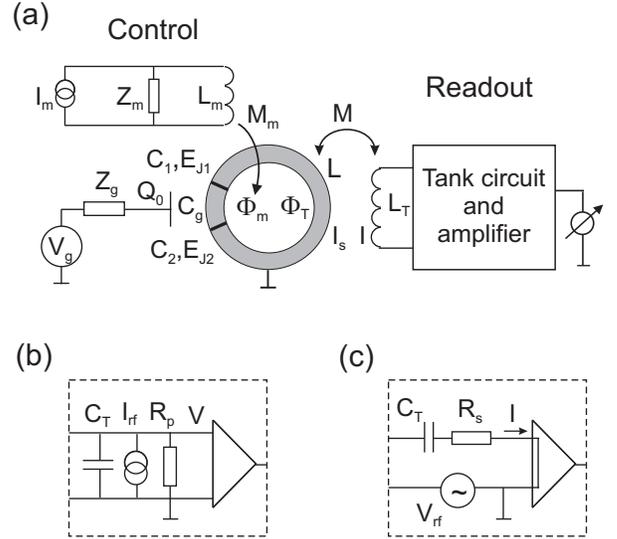}
\caption {(a) The electric circuit diagram of the charge-flux
qubit inductively coupled to a tank circuit by mutual inductance
$M$. The macroscopic superconducting loop of inductance $L$ is
interrupted by two small Josephson tunnel junctions positioned
close to each other and forming a single charge transistor; the
capacitively coupled gate polarizes the island of this
transistor. The qubit is controlled by charge $Q_0$ generated by
the gate and flux $\Phi_m$ induced by coil $L_m$. The tank
circuit which is either of a parallel (b) or a serial (c) type is
driven by a harmonic signal ($I_{\rm rf}$ or $V_{\rm rf}$,
respectively) of frequency $\omega_{\rm rf} \approx \omega_0$,
the resonant frequency of the uncoupled tank circuit.}
\label{Scheme}
\end{center}
\end{figure}

The persistent current is not, however, the only phase-dependent
quantity characterizing the quantum state of the charge-phase
qubit. Another useful quantity is the Josephson inductance of the
double junction, which can be probed by small rf oscillations
induced in the qubit. Recently, we proposed a transistor
configuration of the Cooper pair box (see Fig.\,\ref{Scheme}), in
which the macroscopic superconducting loop closing the transistor
terminals was inductively coupled to a radio-frequency tank
circuit \cite{Zorin-PhysC}. Similar to the rf-SQUID-based method
of measurement of the Josephson junction impedance \cite{Rifk},
this setup makes it possible to measure the rf impedance (more
specifically, the Josephson inductance) of the system of two
small tunnel junctions connected in series and in doing so, to
probe the macroscopic states of the qubit.

On the one hand, the advantage of this method consists in an
effective decoupling between the qubit and a measurement device,
that reduces the decoherence of the qubit. Moreover, the loop
design of the qubit has a potential to perform a data readout in a
non-destructive way \cite{Averin-QND}. On the other hand, due to
the selective characteristic of the tank, the bandwidth of this
setup is rather narrow, so the optimum relation between the
relaxation time of the qubit and the time of measurement becomes
an issue. Furthermore, the driving rf signal may induce
appreciable frequency modulation and dephasing of the qubit
during its evolution (performance of the quantum operations).
Switching the oscillations off and on is, however, possible only
on a relatively long-time scale of a transient process in the
tank.

In this paper we address the problem of decoherence induced in
the charge-flux qubit by the classical resonance tank circuit.
Besides, we propose a strategy of measurement and optimize the
regime of qubit operation for typical parameters of the circuit.

\section{Background}

The small tunnel junctions of the charge-flux qubit are
characterized by self-capacitances $C_1$ and $C_2$ and the
Josephson coupling strengths $E_{J1}$ and $E_{J2}$. These
junctions with a small central island in-between and a
capacitively coupled gate therefore form a single charge
transistor connected in our network as the Cooper pair box (see
Fig.\,\ref{Scheme}). Critical currents of the junctions are equal
to $I_{c1,c2}=\frac{2\pi}{\Phi_0} E_{J1,J2}$, where $\Phi_0=h/2e$
is the flux quantum, and their mean value
$I_{c0}=\frac{1}{2}(I_{c1}+I_{c2})$. The design enables magnetic
control of the Josephson coupling in the box in a dc SQUID
manner. The system therefore has two parameters, the total
Josephson phase across the two junctions $\phi =
\varphi_1+\varphi_2= 2\pi \Phi/\Phi_0$ controlled by flux $\Phi$
threading the loop and the gate charge $Q_0$ set by the gate
voltage $V_g$. The geometrical inductance of loop $L$ is assumed
to be much smaller than the Josephson inductance of the junctions
$L_{J0}=\Phi_0/(2\pi I_{c0})$,
\begin{equation}
\label{beta-L} \beta_L= L/L_{J0} \ll 1.
\end{equation}

Neglecting the magnetic energy term associated with the current
through the small inductance $L$, the Hamiltonian of the
autonomous qubit circuit is expressed as
\begin{equation}
\label{H0} H_0  = \frac{(2en-Q_0)^2}{2C} - E_J(\phi )\cos \chi.
\end{equation}
The second term in Eq.\,(\ref{H0}) originates from the total
Josephson energy equal to $-E_{J1}\cos\varphi_1-E_{J2}\cos
\varphi_2$. The effective Josephson coupling strength is
\begin{equation}
\label{EJ} E_J(\phi) = \left( {E_{J1}^2 + E_{J2}^2 + 2E_{J1}
E_{J2} \cos \phi } \right)^{1/2},
\end{equation}
$|E_{J1}-E_{J2}|\leq E_J(\phi)\leq E_{J1}+E_{J2} \equiv 2E_{J0} =
\frac{\Phi_0}{\pi}I_{c0}$, while a phase variable
$\chi=\varphi+\gamma(\phi)$. The angle $\gamma$ is given by the
expression
\begin{equation}
\label {gamma} \tan \gamma = (j_1-j_2) \tan(\phi/2),
\end{equation}
where dimensionless Josephson energies
$j_{1,2}=E_{J1,J2}/(2E_{J0})$ with $j_1+j_2=1$. The difference
phase $\varphi=\frac{1}{2}(\varphi_1-\varphi_2)$ is a variable
conjugate to the island charge $2en=-2ei
\frac{\partial}{\partial\varphi}$ and $n$ is the operator of the
number of excess Cooper pairs on the island. This charge enters
the charging energy (first) term in Eq.\,(\ref{H0}), in which $C$
is the total capacitance of the island, $C=C_1+C_2+C_g\approx
C_1+C_2$ and the gate capacitance $C_g\ll C_{1,2}$. The
characteristic charging energy, $E_c=e^2/2C$ is assumed to be of
the order of the Josephson coupling energies $E_{J1}\sim E_{J2}\gg
k_BT$.

\begin{figure}[b]
\begin{center}
\leavevmode
\includegraphics[width=2.8in]{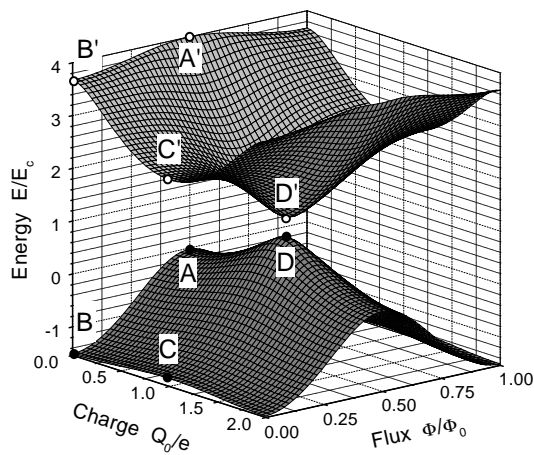}
\caption {Shape of energy bands $E_0$ and $E_1$ in the charge-flux
qubit calculated for the mean Josephson coupling $E_{J0}\equiv
\frac{1}{2}(E_{J1}+E_{J2})=2E_c$ and the Josephson coupling
asymmetry parameter
$|j_1-j_2|=(E_{J1}-E_{J2})/(E_{J1}+E_{J2})=0.1$. Black (hollow)
circles on the zero (excited) band surface mark the locations of
magic points $A \:(A'), B  \:(B'),$ and $C \:(C')$ and the energy
level avoided crossing point $D \:(D')$.} \label{Eigenenergies}
\end{center}
\end{figure}

The Schr\"{o}dinger equation corresponding to the Hamiltonian
Eq.\,(\ref{H0}) is the Mathieu's equation \cite{Abr-Steg}. The
eigen energies form the Bloch bands and eigenfunctions $|n,
q\rangle$ are the Bloch wave functions of a particle in the
periodic (Josephson) potential with "quasimomentum" (here rather
quasicharge) $q$. Its value is the charge which the gate source
provides to the island, i.e., $q=Q_0=C_g V_g$. Each of such eigen
functions can be presented as a coherent superposition of the
plain waves,
\begin{equation} \label{WaveFunc}
|q, n \rangle=\sum\limits_m
C_m^{(n)}\exp\left[i\left(\frac{q}{2e}+m\right)\chi\right],
\end{equation}
where $m = 0, \pm 1, \pm 2, ...$ is the number of excess Cooper
pairs on the island \cite{AZL, LiZo}. The weights of these
coherent contributions $|C_m^{(n)}|^2$ depend on $q$, the band
index $n$ and the characteristic ratio
\begin{equation} \label{Lambda}
\lambda=E_J(\phi)/E_c.
\end{equation}
The two lowest energy levels $E_n(q,\phi)$, i.e., $n=0$ and $1$
(see their dependencies on $q$ and $\phi$ in
Fig.\,\ref{Eigenenergies}), form the basis $\{|0\rangle,
|1\rangle\}$ suitable for qubit operation. In this basis, the
Hamiltonian (\ref{H0}) is diagonal,
\begin{equation} \label{Ho-sigma} H_0=\frac{1}{2}\epsilon\sigma_z,
\end{equation}
where $\sigma_i$ with $i=x,y,z$ is Pauli spin operator. The
general state of qubit is
\begin{equation} \label{General-state} |\Psi\rangle = a|0\rangle
+ b|1\rangle,
\end{equation}
with $|a|^2+|b|^2=1$. It is remarkable that the level spacing
$\epsilon (q,\phi)\equiv \hbar\Omega=E_1(q,\phi)-E_0(q,\phi)$ and,
therefore, the transition frequency $\Omega$ are efficiently
controlled by two knobs, i.e., by varying the parameters $q$ and
$\phi$ (or, equivalently, $Q_0$ and $\Phi$) \cite{Quantr}.

The idea underlying the measurement of this charge-flux qubit is
based on inducing radio frequency oscillations in the tank circuit
of frequency $\omega_{\rm rf} \ll \Omega$ \cite{Zorin-PhysC}. Due
to inductive coupling $M$, these oscillations cause oscillations
of corresponding flux $\Phi_T$ (see Fig.\,1a) and, therefore, of
total phase,
\begin{equation} \label{phase_osc} \phi =\frac{2\pi}{\Phi_0}(\Phi_T+\Phi_m)=
\phi_a \sin (\omega_{\rm rf} t + \theta) + \phi_0.
\end{equation}
If the rf drive signal is sufficiently weak, the amplitude
$\phi_a$ of these oscillations is relatively small, $\phi_a \ll
\pi$. In this linear regime the reverse Josephson inductance is
equal to
\begin{equation} \label{L_J} L_{J}^{-1}(n,q,\phi) =
\left(\frac{2\pi}{\Phi_0} \right)^2 \frac{\partial^2
E_n(q,\phi)}{\partial \phi^2},
\end{equation}
i.e. it is determined by the local curvature of the energy surface
$E_n$. For example, for $E_{J0}= 2E_c$ (see Fig.\,2) at $q\approx
0$, we have the following estimates within the zero and first
bands:
\begin{equation}
\label{LJ-value0} L_{J}^{-1}(0,0,\phi) \approx 0.4 L_{J0}^{-1}
\cos\phi,
\end{equation}
and
\begin{equation}
\label{LJ-value1} L_{J}^{-1}(1,0,\phi) \approx 0.1 L_{J0}^{-1}
\cos\phi,
\end{equation}
respectively. In the vicinity of the level avoided crossing point,
$q=e \textrm{ and } \phi=\pi$ (in Fig.\,2 marked as $D$-$D'$), the
reverse inductances may increase significantly,
\begin{equation}
\label{L_J-value-e} L_{J}^{-1}(n,0,\pi) \approx
\frac{(-1)^{n+1}}{4 |j_1-j_2|}L_{J0}^{-1}, \quad n=0 \textrm{ and
} 1,
\end{equation}
due to small asymmetry of the transistor parameters, $|j_1-j_2|\ll
1$. For example, in the case presented in Fig.\,2, $|j_1-j_2|=0.1$
and $L_{J}^{-1}=\mp 2.5 L_{J0}^{-1}$ for the zero and first band,
respectively. In points $C$ and $C'$, the absolute values
$|L_{J}^{-1}|$ are smaller but the signs for $n=0$ and 1 are still
different.

Coupling to the qubit causes a shift of the resonance frequency
$\omega_0=(L_T C_T)^{-1/2}$ of the tank circuit, i.e.,
$\omega'_0(n)= \omega_0+ \delta\omega_0(n)$, where
\begin{equation} \label{ksi} \delta\omega_0(n)=-\frac{1}{2}k^2\beta_L
\frac{L_{J0}}{L_J(n,q,\phi)}\omega_0.
\end{equation}
Here $k$ is the dimensionless coupling coefficient,
\begin{equation} \label{k-coupl} k=\frac{M}{\sqrt{L_T L}}<1.
\end{equation}
The resonance frequency shift $\delta\omega_0(n)$ carrying
information about the qubit state $|n\rangle$ is found from the
amplitude or/and phase of forced oscillations in the tank. For
achieving sufficient resolution in such measurements, the quality
factor of the tank circuit $Q$ should be about, or larger than,
the ratio $\omega_0/|\delta\omega_0(0)-\delta\omega_0(1)|$.

\section{Inherent and external sources of decoherence}

In our consideration we had neglected the quasiparticle tunneling
which inevitably causes dissipation of energy. Even rare tunneling
of individual quasiparticles across the tunnel junctions, i.e., on
and from the island, can decohere the qubit and completely destroy
the readout regime described above. These processes lead to sudden
change of the operation point, $q \rightarrow q\pm e$ and,
possibly, the energy band index, i.e., cause relaxation $1
\rightarrow 0$.

The processes of single quasiparticle tunneling across a small
Josephson junction had been studied by Averin and Likharev in
Refs.\,\cite{Averin-FNT, AvLikh}. They had generalized the
orthodox theory of single electron tunneling to the case of finite
Josephson coupling, $E_J \neq 0$, taking into account the dynamics
of essential phase factors $\exp(\pm i\chi/2)$ in the electron
tunneling terms added to the Hamiltonian of type  Eq.\,(\ref{H0}).
These factors are the operators of single-electron transfer and
their nonzero matrix elements in our basis are
\begin{equation}
\label{exp-e2-01} e_{nn'}^{\pm} = \langle n, q|\exp(\pm
i\chi/2)|q\pm e, n'\rangle.
\end{equation}
The rates of transitions $|q, n\rangle \rightarrow |q\pm e,
n'\rangle$ are given by
\begin{equation}
\label{Gamma-nn'} \Gamma_{nn'}^{\pm} = |e_{nn'}^{\pm}|^2
\frac{I_{\rm qp}(\epsilon_{nn'}^\pm/e)}{e} \left[1 - \exp \left(-
\frac{\epsilon_{nn'}^\pm}{k_BT} \right)\right]^{-1}.
\end{equation}
In our case $I_{\rm qp}(U)$ is the quasiparticle current-voltage
dependence of the network of two tunnel junctions of the qubit
connected in parallel. Due to 2$e$-periodicity of the energy
surfaces, the corresponding energy gains are identical,
\begin{equation}
\label{E-gain}
\epsilon_{nn'}^+=\epsilon_{nn'}^-=E_n(q,\phi)-E_{n'}(q\pm e,\phi),
\end{equation}
and their value depends on the operation point $\{Q_0, \Phi\}$
(see Fig.\,2).

\begin{figure}[t]
\begin{center}
\leavevmode
\includegraphics[width=3.0in]{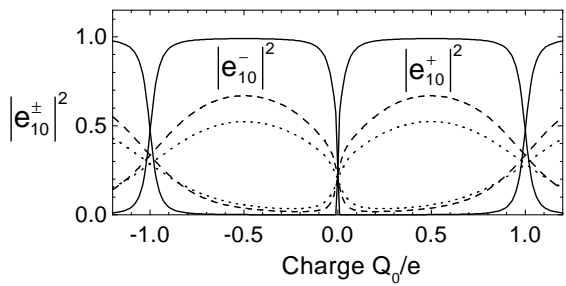}
\caption {Off-diagonal matrix elements of the single quasiparticle
transfer operators $\exp(\pm i\chi/2)$ computed for different
values of equivalent Josephson coupling set by flux $\Phi =
\Phi_0/2$ (solid lines), $\Phi_0/4$ (dashed lines) and $0$ (dotted
lines). The qubit parameters are the same as in Fig.\,2.}
\label{e2ME}
\end{center}
\end{figure}

The relation between this energy and the superconductor energy gap
$\Delta_{\rm sc}$ is important for having the quasiparticle
transitions infrequent or even get rid of them. First, if voltage
$U=\epsilon_{nn'}^\pm/e \leq 2\Delta_{\rm sc}/e$, the
quasiparticle current $I_{qp}(U)$ entering Eq.\,(\ref{Gamma-nn'})
is exponentially small, i.e., $\sim I_{c0}\exp(-\Delta_{\rm
sc}/k_BT)$ \cite{Ramo}. At larger voltages, $U>2\Delta_{\rm
sc}/e$, current $I_{qp}$ is enormously large, $\geq 2I_{c0}$.
Therefore, in order to prevent intensive tunneling of
quasiparticles the gain of energy, $\epsilon_{nn'}^\pm$, should
never exceed $2\Delta_{\rm sc}$. Secondly, if this gain is
smaller than $\Delta_{\rm sc}$, then infrequent quasiparticle
tunneling can, in principle, be blocked by the gap energy
associated with one unpaired electron in the superconducting
island (the so-called even-odd parity effect) \cite{Tuominen}.

Suppression of quasiparticle transitions within the zero energy
band in superconducting Al single charge transistors and Cooper
pair boxes was extensively investigated experimentally. Depending
on experimental skill and luck (see, e.g., \cite{Joyez, Amar,
Mann, Duty}) the inspected devices often exhibited pure Cooper
pair behavior when their charging energy $E_c$ was not larger
than $\sim 100\,\mu$eV $\approx 0.5\Delta_{\rm Al}$, where
$\Delta_{\rm Al}$ is superconductor energy gap of aluminium.
Since the energy gain for transitions in the Cooper pair boxes and
low-voltage-biased transistors, $\epsilon_{00}^\pm$, is less than
$E_c$ for any $E_J$, the condition $E_c<\Delta_{\rm sc}$ can in a
"good" qubit sample ensure suppression of quasiparticle tunneling
in the ground state.

For quasiparticle transitions from the excited state this
condition is clearly insufficient. For example, for small $E_J$
(corresponding to flux value $\Phi=\Phi_0/2$, Eq.\,(\ref{EJ}))
the energy gain values are between about $E_c$ (for the process
$D'\rightarrow A$, see Fig.\,2) and $4E_c$ (for the processes
$A'\rightarrow D'$ and $A'\rightarrow D$). At larger $E_J$ both
the minimum and maximum energy gain values are even larger. For
example, for $E_J=4E_c$ (i.e., $\Phi=0$), transitions
$C'\rightarrow B$ and $B'\rightarrow C$ correspond to energy
$\approx 4E_c$ and $\approx 5E_c$, respectively. Since the first
multiplier factor in the expression for resulting relaxation rate,
\begin{equation}
\label{Gamma-qp-sum} [\tau_r^{\rm (qp)}]^{-1} = \Gamma_{10}^+
+\Gamma_{10}^- \approx (|e_{10}^{+}|^2+|e_{10}^{-}|^2)
\frac{I_{\rm qp}(\epsilon_{10}^\pm /e)}{e},
\end{equation}
is nonzero for any $Q_0$ and $\Phi$ (see the plots of the two
items in Fig.\,\ref{e2ME}), only the condition $E_c \leq
\Delta_{\rm sc}/5$ can ensure suppression of these transitions in
arbitrary operation point of our qubit. Possibly insufficiently
small value of $E_c$ was the reason of very short relaxation time
(tens of ns) in the recent experiment with a charge qubit by Duty
et al. \cite{Duty}. Their Al Cooper pair box had $E_c \approx 0.8
\Delta_{\rm sc}$ and $E_J \approx 0.4E_c$, so the energy gain in
the chosen operation point ($Q_0=0.4e$) was too large, i.e.,
about $2.2E_c \approx 1.8 \Delta_{\rm sc} > \Delta_{\rm sc}$
(although in the ground state this sample nicely showed the pure
Cooper pair characteristic).

Moreover, there are several sources of decoherence due to
coupling of the qubit to the environmental degrees of freedom. For
evaluating the effect of these sources on the qubit the coupling
Hamiltonian term, $H_{\rm coupl}=H_c^{(e)}+H_c^{(m)}$, is
included in the total Hamiltonian of the system,
\begin{equation} \label{Ham-total} H=H_0+H_{\rm coupl}+H_{\rm
bath},
\end{equation}
where $H_{\rm bath}$ is a bath operator. $H_c^{(e)}$ and
$H_c^{(m)}$ are the electric control line term and the magnetic
coupling term respectively. The latter is associated with both the
flux control line and the tank circuit. Fluctuations originating
from the sources of gate- and flux-control lines may, in
principle, lead to significant decoherence of the qubit. As was
shown in Ref.\,\cite{Makhlin} and demonstrated in experiments
\cite{Nakamura, CTH-rf-SET, Vion}, these effects can, however, be
minimized by choosing appropriate (minimum) coupling. On the
other hand, decoherence caused by the tank-circuit-based readout
system requires special analysis, because weakening of this
coupling leads the input signal being reduced. Below we will
start with sources of decoherence associated with the control
lines and then will analyze the effect of the tank circuit and
amplifier.

\section{Coupling to charge control line}

The coupling of the charge-phase qubit to the electrical control
line is actually similar to that of the gate coupling in the
ordinary Cooper pair box \cite{Makhlin}. However, here we assume
that Josephson coupling parameter $\lambda$ is not necessarily
small, as is usually assumed in the analysis of the charge
qubits. This generalization of the model is essential because the
external flux $\Phi_m$ changes the effective Josephson energy
Eq.\,(\ref{EJ}) of the qubit over a wide range. The assumption of
not small $\lambda$ implies that the eigenstates of our system
Eq.\,(\ref{WaveFunc}) are generally composed of several (not only
two) plane-wave states.

The coupling term can be presented as
\begin{equation} \label{Hc-e} H_{c}^{(e)} =-2en \delta V_e,
\end{equation}
where $\delta V_e$ is the operator of voltage fluctuations on the
island in the absence of Josephson coupling. The charge operator
is equal to $2en = Q_0-C\hat{V}$, so the essential part of the
coupling Hamitonian is
\begin{equation} \label{Hc1} H_{c}^{(e)} = C \hat{V} \delta V_e.
\end{equation}
The voltage operator is equal to
\begin{equation} \label{V-hat} \hat{V}=\frac{\Phi_0}{2\pi} \,\dot{\varphi}
=\frac{\Phi_0}{2\pi} \left( \dot{\chi} -
\frac{\partial\gamma}{\partial\phi}\, \dot{\phi}\right)
=\frac{\Phi_0}{2\pi} \,\dot{\chi}.
\end{equation}
Here we assumed slow variation of the total phase $\phi$,
Eq.\,(\ref{phase_osc}). The operator of voltage $\hat{V}$ is
similar to the operator of velocity of an electron in the periodic
electric potential of the crystal lattice \cite{Lifshitz}, and
its interband matrix elements are
\begin{equation} \label{V_repr}
V_{nn'} =\frac{\partial E_n}{\partial q}
\delta_{n,n'}+i\frac{E_n-E_{n'}}{2e}\chi_{nn'}(1-\delta_{n,n'}),
\end{equation}
where $\delta_{n,n'}$ is the Kroneker delta and $\chi_{nn'}$ are
the matrix elements of the phase operator $\chi$ \cite{LiZo}.

\begin{figure}[b]
\begin{center}
\leavevmode
\includegraphics[width=2.8in]{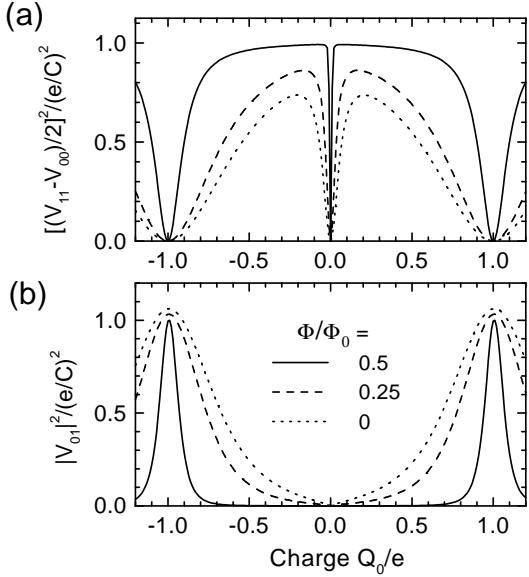}
\caption {The terms composed of diagonal (a) and off-diagonal (b)
matrix elements of operator $\hat{V}$ and entering
Eqs.~(\ref{Vmod}) and (\ref{Vtan}) are presented for different
values of flux $\Phi_e$ for the given qubit parameters (see
caption of Fig.\,2).} \label{V01V0011}
\end{center}
\end{figure}

Finally, the coupling Hamiltonian, Eq.\,(\ref{Hc1}), takes the
form
\begin{equation} \label{Hc-sigma} H_c^{(e)} = ( \sigma_x \sin
\eta_e +\sigma_z \cos\eta_e ) X_e,
\end{equation}
where operator
\begin{equation} \label{Xe} X_e = C \|V\| \delta V_e
\end{equation}
with
\begin{equation} \label{Vmod} \|V\| = \frac{1}{2}\sqrt{(V_{11}-V_{00})^2
+4|V_{01}|^2}
\end{equation}
and
\begin{equation}
\label{Vtan} \tan\eta_e= \frac{2|V_{01}|}{(V_{11}-V_{00})}.
\end{equation}
(The plots of the terms entering Eqs.\,(\ref{Vmod}) and
(\ref{Vtan}) obtained by numerical calculations are presented in
Fig.\,\ref{V01V0011}.) Thus, $X_e = \sum_a C_a x_a$ can be
considered as an operator of the bath \cite{Cald-Legg} with the
Hamiltonian
\begin{equation} \label{H_e} H_b^{(e)} =
\sum_a \left( \frac{p_a^2}{2m_a}+\frac{m_a \omega_a^2 x_a^2}{2}
\right)
\end{equation}
and the spectral density $S_X^{(e)}(\omega)=C^2 \|V\|^2
S_{V}^{(e)}(\omega) = J_e(\omega)\Theta(\omega,T)/\omega$. Here
the oscillator energy function is
\begin{equation}
\Theta(\omega,T)=\frac{\hbar
\omega}{2}\coth\frac{\hbar\omega}{2k_B T}
\end{equation}
and
\begin{equation} \label{J} J_e(\omega)= \frac{\pi}{2} \sum_a
\frac{C_a^2}{m_a \omega_a} \delta(\omega-\omega_a).
\end{equation}
Assuming smallness of $C_g$, the spectral density $S_{V}^{(e)}$ of
the fluctuations of  $\delta V_e$ is given by
\begin{equation} \label{Sv} S_{V}^{(e)}(\omega) =
\frac{2}{\pi}\left(\frac{C_g}{C}\right)^2 {\rm Re}Z_t(\omega)\,
\Theta(\omega,T),
\end{equation}
where $Z_t=(i\omega C_g +Z_g^{-1})^{-1}$ is determined by parallel
connection of the qubit capacitance $C_gC/(C_g+C)\approx C_g$ and
the gate line impedance $Z_g(\omega)\sim R_{100} \equiv
100\,\Omega$. Therefore, for frequencies up to $\omega_g\equiv
(R_{100}C_g)^{-1} \gg \epsilon/\hbar$, i.e., at all characteristic
frequencies of the system, ${\rm Re}Z_t=R_{100}$. This is the case
of linear damping in the Caldeira-Leggett model,
\begin{equation} \label{Jlin} J_e(\omega)= \frac{\pi}{2} \alpha_e
\hbar\omega,
\end{equation}
with the dimensionless factor
\begin{equation} \label{alpha} \alpha_e(q,\phi)=\left(\frac{C_g \|V\|}{e}\right)^2
\frac{R_{100}}{R_Q} \lesssim \left(\frac{C_g}{C}\right)^2
\frac{R_{100}}{R_Q},
\end{equation}
where $R_Q=h/4e^2\approx 6.45\, {\rm k}\Omega$, the resistance
quantum. The estimate similar to the last expression in
Eq.\,(\ref{alpha}), was given earlier in Ref. \cite{Makhlin} for
small $\lambda$.

Relaxation and dephasing caused by the charge control line can,
therefore, be described by the spin-boson model with linear
damping \cite{Legg}. The corresponding rates are given by the
expressions
\begin{equation} \label{tau-e} [\tau_{r}^{(e)}]^{-1} =
 \pi \alpha_e \sin ^2\eta_e\,
\Omega \,\coth \frac{\hbar\Omega}{2k_B T},
\end{equation}
and
\begin{equation} [\tau_{\varphi}^{(e)}]^{-1} = [2 \tau_{r}^{(e)}]^{-1}
+ \pi \alpha_e \cos^2 \eta_e \,\frac{2k_BT}{\hbar}.
\end{equation}
One can see that in accordance with the conclusions of
Refs.\,\cite{Makhlin,Wal}, reducing the coupling coefficient
$\alpha_e$ by a small factor $(C_g/C)^2\ll 1$ can significantly
depress the decoherence rates.

\section{Coupling to flux control line}

The inductive coupling of the qubit loop with the control and
readout circuits is described by the Hamiltonian
\begin{equation} \label{Hcoupl} H_c^{(m)} =-\hat{I}_s
(\delta\Phi_m+\delta\Phi_T),
\end{equation}
where $\hat{I}_s$ is the operator of current circulating in the
qubit loop, $\delta\Phi_m = M_m \delta I_m$, the bath operator
(proportional to fluctuations of current $\delta I_m$ in the
control inductance $L_m$); $\delta\Phi_T = M \delta I$ is the
operator of flux associated with current fluctuations in the tank
circuit.

\begin{figure}[b]
\begin{center}
\leavevmode
\includegraphics[width=2.8in]{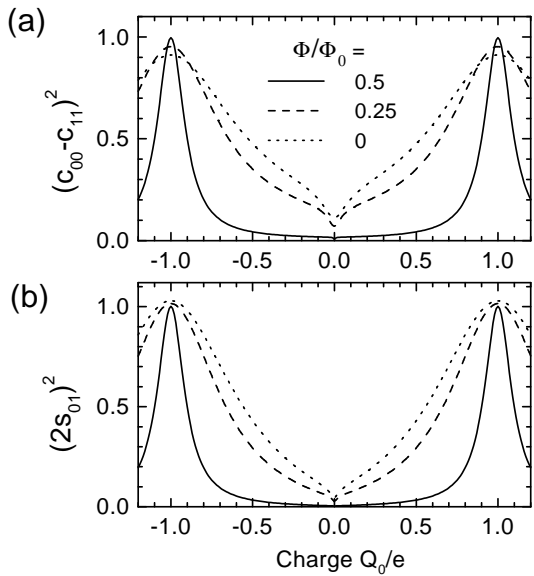}
\caption {Terms composed of diagonal (a) and off-diagonal (b)
matrix elements of operators $\cos\chi$ and $\sin\chi$,
respectively, calculated for different values of flux $\Phi_e$ for
the given qubit parameters (see caption of Fig.~2).}
\label{s01c0011}
\end{center}
\end{figure}

To specify the coupling, we present the operator $\hat{I}_s$ in
the eigen basis (\ref{WaveFunc}), i.e., we find the matrix
elements
\begin{equation} \label{nIsn} \langle n|\hat{I}_s|n'\rangle,\qquad
n,n'=0,\:1.
\end{equation}
In the general case, $\hat{I}_s$ is given by the expression
\begin{equation} \label{Is} \hat{I}_s=
\kappa_1\dot{Q}_1+\kappa_2\dot{Q}_2,
\end{equation}
where the dimensionless factors $\kappa_{1,2}=C_{2,1}/C$ with the
identity $\kappa_1+\kappa_2=1$. Quantities
$Q_{1,2}=-2ei\frac{\partial}{\partial\varphi_{1,2}}$ are the
charges on the first and second junction, respectively, and their
time derivatives are the Josephson supercurrents,
\begin{equation} \label{dotQ}
\dot{Q}_{1,2}=\frac{i}{\hbar}[Q_{1,2},H_0]=I_{c1,c2}\sin\varphi_{1,2}.
\end{equation}
Using identity $\varphi_{1,2}=
\frac{\phi}{2}\pm\varphi=\frac{\phi}{2}\pm\chi\mp\gamma$ and
Eq.\,(\ref{gamma}), the circulating current can be presented as
\begin{equation} \label{IsMod} \hat{I}_s=
I_1(\phi) \cos\chi +I_2(\phi) \sin\chi.
\end{equation}
The amplitudes of these two components are
\begin{equation}
 \label{I-1}
I_1= \frac{2\pi}{\Phi_0} \frac{E_{J1}E_{J2}}{E_J(\phi)}\sin\phi,
\end{equation}
and
\begin{eqnarray}
 \label{I-2} I_2 &&=  (j_1-j_2)(\kappa_1 j_1+\kappa_2 j_2)
 \frac{8\pi}{\Phi_0}\frac{E_{J0}^2}
 {E_J(\phi)}\nonumber\\
 && +(\kappa_1-\kappa_2)\frac{4\pi}{\Phi_0}\frac{E_{J1}E_{J2}}
 {E_J(\phi)}\cos^2 \frac{\phi}{2},
\end{eqnarray}
respectively.

Since the Hamiltonian, Eq.\,(\ref{H0}), is an even function of
$\chi$, the operators $\cos\chi$ and $\sin\chi$ entering
Eq.\,(\ref{IsMod}) are diagonal and off-diagonal, respectively.
Amplitude $I_1$ is merely the classical Josephson current across
two large-capacitance junctions as a function of the overall phase
difference $\phi$, while the diagonal term $\cos \chi$ describes
the suppression of this current due to the charging effect
($E_c\neq0$) (see, for example, Ref.\,\cite{Zorin-prl1}). The
second, off-diagonal, term in Eq.\,(\ref{IsMod}) is due to
asymmetry of the transistor; it gives rise to the interband
transitions $0\leftrightarrow1$. Denoting the nonzero values of
the corresponding matrix elements as
\begin{equation}\label{s-c1}
c_{00}= \langle 0|\cos \chi|0\rangle, \quad c_{11}= \langle 1|\cos
\chi|1\rangle,
\end{equation}
and
\begin{equation}\label{s-c2}
s_{01} = |\langle 0|\sin \chi|1\rangle|,
\end{equation}
we obtain the coupling Hamiltonian Eq.\,(\ref{Hcoupl}) in the form
\begin{equation} \label{Hm-sigma} H_c^{(m)} = ( \sigma_y \sin
\eta_m +\sigma_z \cos\eta_m ) (X_m+X_T),
\end{equation}
where
\begin{equation} \label{XmT} X_{m,T} = -\|I\|\, \delta
\Phi_{m,T},
\end{equation}
\begin{equation} \label{III} \|I\| =
\sqrt{[(c_{11}-c_{00}) I_1]^2+[2s_{01} I_2]^2},
\end{equation}
\begin{equation}
\label{tan-n} \tan\eta_m = \frac{2 s_{01} I_2}{(c_{11}-c_{00})
I_1},
\end{equation}
(see the plots of terms entering these expressions in
Fig.\,\ref{s01c0011}).

First we omit in Eq.\,(\ref{Hm-sigma}) the term $X_T$ associated
with fluctuations of the tank circuit and focus on the effect of
fluctuations in the flux control line $\delta \Phi_m=M_m \delta
I_m$. Assuming real impedance of the flux control line, $Z_m\sim
R_{100}$, we obtain the spectral density of the operator $X_m
\propto \delta I_m$ in the form $S_X^{(m)}(\omega)= M_m^2 \|I\|^2
S_{I}^{(m)}(\omega) = J_m(\omega)\Theta(\omega,T)/\omega$. At
frequencies below $\omega_m \equiv R_{100}/L_m$ the function
$J_m$ is linear,
\begin{equation} \label{J_m} J_m(\omega)= \frac{\pi}{2}
\alpha_m \hbar\omega
\end{equation}
with the dimensionless coupling factor
\begin{equation} \label{alpha_m} \alpha_m(q,\phi)=
\left(\frac{2M_m \|I\|}{\Phi_0}\right)^2 \frac{R_Q}{R_{100}}.
\end{equation}
At higher frequencies, $\omega > \omega_m$, the effective damping
decays as $(\omega_m/\omega)^2$.

In fact, Eq.\,(\ref{alpha_m}) describes the effect of coupling to
the control flux in the general case. An estimate of the coupling
factor based on the evaluation
\begin{equation}
\label{alpha_m_estim} \|I\|\approx \frac{1}{2}\left|\frac{\partial
E_J}{\partial \Phi}\right|,
\end{equation} which is valid for symmetric transistor
($I_2=0$) with small $E_J$, was made by Makhlin et al.
\cite{Makhlin}. Small mutual inductance $M_m$ \cite{Makhlin,Wal}
leads to small $\alpha_m$ and, therefore, causes significant
depression of the corresponding relaxation rate,
\begin{equation} \label{tau-m} [\tau_{r}^{(m)}]^{-1} =
\pi \alpha_m \sin ^2\eta_m\, \Omega \,\coth
\frac{\hbar\Omega}{2k_B T},
\end{equation}
and dephasing rate,
\begin{equation}
[\tau_{\phi}^{(m)}]^{-1} = [2 \tau_{r}^{(m)}]^{-1} + \pi \alpha_m
\cos^2 \eta_m \,\frac{2k_BT}{\hbar}.
\end{equation}

So far we have considered the effects of decoherence due to the
charge and flux control lines as two independent effects.
Actually, one has to describe them together using a multibath
model \cite{Makhlin}. If either of these decoherence effects is
small, i.e. the so-called Hamiltonian-dominated regime is
realized, the total rates due to contributions of the two control
lines are given by
\begin{equation}\label{tau-sum1} [\tau_{r}^{(c)}]^{-1} =
[\tau_{r}^{(e)}]^{-1}+[\tau_{r}^{(m)}]^{-1}, \\
\end{equation}
\begin{equation}
 \label{tau-sum2} [\tau_{\varphi}^{(c)}]^{-1} = [2 \tau_{r}^{(c)}]^{-1} +
[\tau_{\varphi}^{(e)}]^{-1}+[\tau_{\varphi}^{(m)}]^{-1}.
\end{equation}
In our model we shall assume that such a regime is realized and,
moreover, the resulting rates, Eqs.\,(\ref{tau-sum1}) and
(\ref{tau-sum2}), can be made negligibly small. Below we will
focus on the effect of the readout circuit, whose coupling
strength has to be optimized.

\section{Decoherence due to readout system}

In contrast to the control lines, coupling to a readout device (in
our case the tank circuit with amplifier) cannot be made
arbitrarily small in order to reduce the decoherence. This
coupling should ensure sufficiently strong signals at the
amplifier input in order to perform a measurement with a
reasonable signal-to-noise ratio on a time scale shorter than that
determined by other factors, namely $\tau_{r}^{(c)}$. Moreover,
without an efficient switch (see possible design of such a switch
in, e.g., Ref.\,\cite{Clarke-switch}), such a coupling may cause
significant dephasing of the qubit during quantum gate
manipulation.

Inductive qubit coupling to the tank circuit is described by the
Hamiltonian Eq.\,(\ref{Hm-sigma}). The spectral density of
fluctuations of the corresponding variable $X_T \propto
\delta\Phi_T = M\delta I$ is expressed as
\begin{equation} \label{SX-T} S_X^{(T)}(\omega)=
\frac{2}{\pi} M^2 \|I\|^2 S_I^{(T)}(\omega)= J_T(\omega)
\frac{\Theta(\omega,T^*)}{\omega},
\end{equation}
where $S_I^{(T)}(\omega)$ is the spectral density of the noise
current $\delta I$ across inductance $L_T$. Since the cold
(superconducting) tank circuit itself presumably has very low
losses, a backaction noise $\delta I$ of the amplifier is
dominating. It is associated with input real impedance, modeled by
$R_p$ or $R_s$ for parallel and serial configurations,
respectively (see Fig.\,1). $T^*$ is the effective temperature
associated with this impedance.

Spectral density $S_I^{(T)}$ and function $J_T(\omega)$ can be
found from a network consideration. Neglecting small detuning
$\delta \omega_0\ll \omega_0$, in the case of the parallel network
(Fig.\,1b), the spectral function $J_T$ is given by the expression
\begin{equation} \label{J-Tp} J_T^{(p)}(\omega)=
\frac{2}{\pi} \alpha_p \hbar \omega \frac{\omega^4_0
}{(\omega^2-\omega^2_0)^2+\omega^2 \omega^2_0 Q^{-2}},\\
\end{equation}
with
\begin{equation}
\alpha_p= \left(\frac{2M \|I\|}{\Phi_0}\right)^2
\frac{R_Q}{R_p}=\frac{k^2\beta_L}{\pi Q}\frac{\|I\|^2}{e\omega_0
I_{c0}},
\end{equation}
where the quality factor $Q=\omega_0 C_T R_p=R_p/\omega_0 L_T$. In
the case of the serial network shown in Fig.\,1c,
\begin{equation} \label{J-Ts} J_T^{(s)}(\omega)=
\frac{2}{\pi} \alpha_s \hbar \omega \frac{\omega^2
\omega^2_0}{(\omega^2-\omega^2_0)^2+\omega^2 \omega^2_0 Q^{-2}},
\end{equation}
with
\begin{equation} \alpha_s= \left(\frac{2M \|I\|}{\Phi_0}\right)^2
\frac{R_Q R_s}{(\omega_0 L)^2}
\end{equation}
and $Q=(\omega_0 C_T R_s)^{-1}=\omega_0 L_T /R_s$.

In contrast to the linear spectral functions for the control
lines, Eqs.\,(\ref{Jlin}) and (\ref{J_m}), the functions given by
Eqs.\,(\ref{J-Tp}) and (\ref{J-Ts}) describe a structured bath,
viz. they both are of a Lorentzian (resonance) shape. A similar
situation emerges, for example, in the case of the flux qubit
with readout using a $C$-shunted dc SQUID \cite{Wal}. The
theoretical analysis of the spin-boson model with a structured
bath was made by Kleff et al. \cite{Kleff} on the basis of the
flow equations. If coupling is, as in our case, weak, only the
high frequency ($\omega\sim \Omega$) and low frequency ($\omega
\rightarrow 0$) behaviors of $J(\omega)$ account for relaxation
and dephasing, respectively \cite{Wal, Grifoni, Tian}.

Since frequency $\Omega$ is typically about tens of GHz and the
distance between qubit and amplifier presumably exceeds the
wavelength, the effective real admittance of the parallel circuit
at these frequencies is equal to $R_{100}^{-1}$ and the impedance
of the serial circuit $\approx R_{100}$. Therefore, the relaxation
rates increase by factors $g_p=R_p/R_{100}\gg 1 $ and
$g_s=R_{100}/R_s \gg 1$, respectively.

For the parallel tank circuit the relaxation and dephasing rates
(presumably, $\ll \omega_0$) are equal to
\begin{equation} \label{tau-p} [\tau_{r}^{(p)}]^{-1} =
\pi g_p\alpha_p \sin ^2\eta_m\left(
\frac{\omega_0}{\Omega}\right)^4 \Omega \,\coth
\frac{\hbar\Omega}{2k_B T^*},
\end{equation}
and
\begin{equation}
\label{tau-dph-p} [\tau_{\varphi}^{(p)}]^{-1} = [2
\tau_{r}^{(p)}]^{-1} + \pi \alpha_p \cos^2 \eta_m
\,\frac{2k_BT^*}{\hbar},
\end{equation}
respectively. The relaxation rate is dramatically suppressed due
to the small frequency ratio, $(\omega_0/\Omega)\ll 1$. For the
serial configuration, the  corresponding rates are
\begin{equation} \label{tau-s} [\tau_{r}^{(s)}]^{-1} =
\pi g_s \alpha_s \sin ^2\eta_m\left(
\frac{\omega_0}{\Omega}\right)^2 \Omega\, \coth
\frac{\hbar\Omega}{2k_B T^*},
\end{equation}
\begin{equation}
[\tau_{\varphi}^{(s)}]^{-1} = [2 \tau_{r}^{(s)}]^{-1}.
\end{equation}
The dephasing rate is determined by the rate of relaxation,
because at low frequency, $\omega \ll \omega_0$, function
$J_T^{(s)}(\omega) \propto \omega^3$  \cite{Legg}. Due to weaker
decay of the serial circuit impedance at high frequencies,
$\omega \gg \omega_0$, the relaxation rate is, however,
substantially higher than that in the case of the parallel
circuit configuration. We shall therefore focus our further
consideration only on the parallel tank circuit as the more
favorable (allowing longer measuring time).

\section{Magic points and some estimations}

The analysis of the coupling between the qubit and the tank
circuit, Eqs.\,(\ref{Hm-sigma})-(\ref{tan-n}) and Fig.\,(5), shows
that its strength $X_T \propto\|I\|$ and mixing angle $\eta_m$
can be significantly varied by choosing an appropriate operation
point. For example, as can be seen from Eq.\,(\ref{I-1}), the
diagonal component of $X_T$ ($\propto I_1$) which essentially
cause pure dephasing of the qubit, is zero, i.e., mixing angle
$\eta_m=\pi/2$, at phase values $\phi=0$ and $\pi$. The
derivatives $\frac{\partial E_{0,1}}{\partial \phi}$ and,
therefore, the circulating supercurrent are zero. Moreover, as
illustrated in Fig.\,5b, if the gate charge $Q_0\approx 0$ (i.e.,
derivatives $\frac{\partial E_{0,1}}{\partial Q_0}=0$), then
$|s_{01}|$ and, hence, $X_T$ are minimum. In particular, at the
bias flux $\Phi_m=\Phi_0/2$ or, equivalently, $\phi=\pi$ (this
point is marked in Fig.\,2 as $A$), $E_J(\phi)=|E_{J1}-E_{J2}|\ll
E_c$, so one can use the explicit expressions for the wave
functions, Eqs.\,(A11) and (A12) of Ref.\,\cite{LiZo}, and obtain
\begin{equation}
|s_{01}|=\frac{1}{16\sqrt{2}}\frac{E_J(\phi)}{E_c}
=\frac{|j_1-j_2|}{8\sqrt{2}}\frac{E_{J0}}{E_c}.
\end{equation}
Then the value of $\|I\|$ (\ref{III}) is
\begin{equation}
\label{I-phi-pi} \|I \|_A =
2|s_{01}|I_2\approx\frac{|j_1-j_2|}{4\sqrt{2}}\frac{E_{J0}}{E_c}I_{c0},
\end{equation}
where we have taken into account that $\kappa_1 \sim \kappa_2 \sim
0.5$ and the second term in Eq.\,(\ref{I-2}), $\propto (\kappa_1-
\kappa_2)$, vanishes due to the relation $\cos (\phi/2)=0$. In the
point $Q_0=0,\phi=0$ (marked as $B$ in Fig.\,2), the Josephson
energy $E_J(\phi)=2E_{J0}$ and $2|s_{01}|$ is approximately equal
to $(1/8\sqrt{2})E_{J0}/E_c$, so
\begin{equation}
\label{I-phi-0} \|I \|_B
\approx\frac{|j_1-j_2+\kappa_1-\kappa_2|}{8\sqrt{2}}\frac{E_{J0}}{E_c}I_{c0},
\end{equation}
while for $Q_0=e$ (point $C$  in Fig.\,2), $|s_{01}|\approx 0.5$
and
\begin{equation}
\label{I-Q0-e} \|I \|_C \approx |j_1-j_2+\kappa_1-\kappa_2|I_{c0}.
\end{equation}
It is remarkable that the effect of asymmetry in critical currents
and capacitances of the junctions can, in principle, cancel if
$(j_1-j_2)=-(\kappa_1-\kappa_2)$. In practice, however, the signs
of $(j_1-j_2)$ and $(\kappa_1-\kappa_2)$ normally are similar
because the critical current and capacitance are both proportional
to the junction area and such cancelling does not occur.

Comparing Eqs.\,(\ref{I-phi-pi}), (\ref{I-phi-0}) and
(\ref{I-Q0-e}) one can see that on the assumption of small
asymmetry of the transistor, $j_1\approx j_2 \approx \kappa_1
\approx \kappa_2 \approx 0.5$, the coupling strength $\alpha_p$
in points $A\,(Q_0=0,\: \phi=\pi)$, $B\,(Q_0=0,\: \phi=0)$ and
$C\,(Q_0=e,\: \phi=0)$ is rather small, but it is significant in
the point, $D\,(Q_0=e,\: \phi=\pi)$, where parameter
$|s_{01}|\approx 0.5$ and
\begin{equation}
\label{I-Q0-e-pi} \|I \|_D \approx I_{c0}.
\end{equation}
To illustrate this behavior, the coupling strength evaluated for
typical parameters of the system is presented in Table\,1.

From the point of view of operation with a minimum dephasing rate,
the "magic" points $A$, $B$ and $C$, in which supercurrent
$I_1=0$ (see Eq.\,(\ref{I-1})), are clearly preferable to other
points in the $Q_0$-$\Phi$ plane. Therefore, manipulation of the
qubit can, in principle, be performed in any of these points. The
estimated values of the corresponding quality factor for quantum
manipulation, $Q_\varphi\equiv\Omega\tau_{\varphi}^{(p)}$, given
in Table\,1, are sufficiently high. For example, in the case of
preparation of the qubit in point $A$, the manipulation can be
performed by means of a dc pulse applied to the transistor gate
\cite{Nakamura, CTH-rf-SET, Duty}. This pulse (with short rise
and fall times) can rapidly switch the qubit, for example, to
point $D$ and back to $A$ causing its evolution (although with
significant dephasing) during the pulse span. Our qubit prepared
in the ground state in point $A$ or $B$ or $C$ can be
(preferably) manipulated by a pulse of microwave frequency, $\sim
\Omega$, applied to the gate. For example, the Quantronium qubit
in the experiment by Vion et al. \cite{Vion} was manipulated by
microwave pulses in point $C$.

\begin{table}
\caption{Evaluated qubit parameters derived on the assumption
$E_{J0}=2E_c=80\,\mu$eV (i.e., $I_{c0}\approx 40$\,nA and $5E_c =
\Delta_{\rm Al} \approx 200\,\mu$eV, the energy gap of Al) and
$j_1-j_2=\kappa_1-\kappa_2=0.1$. The tank circuit quality factor
$Q=100$, frequency $\omega_0=2\pi\times 100$\,MHz,
$(L_T/C_T)^{1/2}=100\,\Omega$, $k^2Q\beta_L=20$ and temperature
$T^*=1$\,K $\gg T\sim 20$\,mK. As long as the dephasing rate in
the magic points is nominally zero, a 0.1\% inaccuracy of the
adjustment of the values $\phi=\pi$ and $0$ was assumed.}
\label{table1}
\begin{ruledtabular}
\begin{tabular}{ccccc}
Operation point: &$A$-$A'$&$B$-$B'$&$C$-$C'$&$D$-$D'$\\
\tableline Frequency ${\Omega}/2\pi$ \ [GHz]& $39$&$50$&$36$&$4$\\
Coupling strength $\alpha_p$ &
$2\,10^{-2}$&$10^{-2}$&$4\,10^{-2}$&$1$\\
Qubit quality factor $Q_{\varphi}$
& $3\,10^{4}$&$2\,10^{5}$&$10^{4}$&$<30$\\
Relaxation time $\tau_{r}^{(p)}$ \ [s]&
$8\,10^{-2}$&$10^{-1}$&$6\,10^{-3}$&$10^{-7}$
\end{tabular}
\end{ruledtabular}
\end{table}

\begin{figure}[t]
\begin{center}
\leavevmode
\includegraphics[width=2.9in]{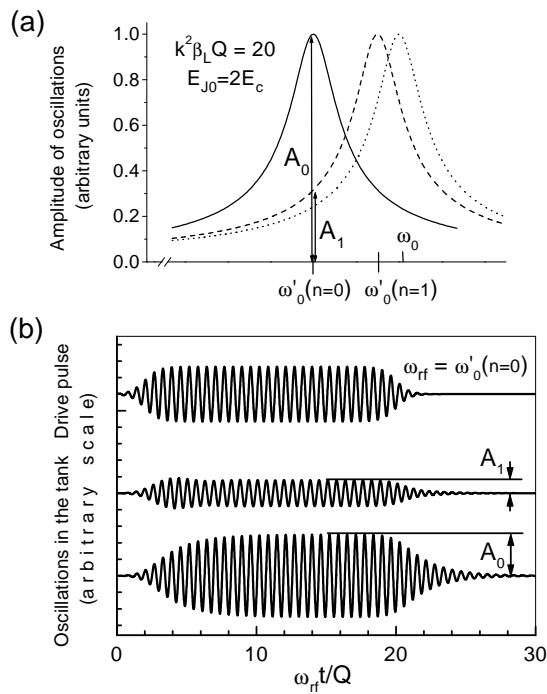}
\caption {The principle of narrow-band radio-frequency readout of
the qubit. (a) The resonance curves of the uncoupled (dotted
line) and coupled to the qubit tank circuit biased at operation
point $A$ in the excited state (dashed line) and in the ground
state (solid line). (b) Driving pulse applied to the tank circuit
(top curve) and the response signal of the tank in resonance (the
ground qubit state, bottom curve) and off-resonance (excited
state, middle curve). A smooth envelope of the driving pulse is
used to suppress transient oscillations and has a small effect on
the rise time of the response signal. For clarity the curves are
shifted vertically.} \label{Pulse}
\end{center}
\end{figure}

For reading out the final state, the qubit dephasing is of minor
importance, while the requirement of a sufficiently long
relaxation time is decisive. Moreover, the relaxation rate may
somewhat increase due to oscillations in the tank induced by a
drive pulse (see Fig.\,\ref{Pulse}), which leads to the
development of oscillations around a magic point along $\phi$
axis, Eq.\,(\ref{phase_osc}). If the frequency of these
oscillations is sufficiently low, $\omega_{\rm rf} \ll \Omega$,
they result only in a slow modulation of transition frequency
$\Omega$. Increase in amplitude of steady oscillations up to
$\phi_a\approx \pi/2$ (determined by the amplitude of the drive
pulse and detuning) yields a large output signal and still
ensures the required resolution in the measurement provided the
product $k^2Q \beta_L>1$ is sufficiently large. (At larger
amplitudes $\phi_a$, the circuit operates in a non-linear regime
probing an averaged reverse inductance of the qubit whose value,
as well as the produced frequency shift $\delta\omega_0$, is
smaller \cite{Zorin-prl2}.) As points $A$ and $B$ lie on the axis
$Q_0=0$ and are both characterized by a sufficiently long
relaxation time, reading-out of the qubit state with the rf
oscillation span $\pm\pi/2$ is preferable in either point. In the
case of operation point $C$, the limited amplitude of the
oscillations does not reduce much the relaxation time either.
Significant reduction of the relaxation time occurs in the
vicinity of point $D$. Due to this property which is due to the
dependence of the transversal coupling strength on $\phi$,
Eqs.\,(\ref{I-2})-(\ref{tan-n}), the measurement of the
Quantronium state using a switching current technique was
possible in the middle of segment $CD$ (see Fig.\,2), where the
maximum values of the circulating current in the excited and
ground states were of different sign \cite{Vion}.

In the vicinity of energy level avoided crossing point $D$, in
which the gap between the zero and first excited states is
minimum, $\hbar\Omega=2|j_1-j_2|E_{J0}$, oscillations of $\phi$
may cause Landau-Zener transitions $|0\rangle \leftrightarrow
|1\rangle$ \cite{LandZen}. Probability of a such transition per
single sweep
\begin{equation}
\label{LZ} p_{\rm LZ} = \exp \left[-2\pi\frac{(j_1-j_2)^2
E_{J0}}{\phi_a\hbar \omega_{\rm rf}} \right],
\end{equation}
can be appreciable in a sufficiently symmetric transistor and/or
at high driving frequency $\omega_{\rm rf}$, i.e., when $|j_1-j_2|
\lesssim(\hbar \omega_{\rm rf}/E_{J0})^{1/2}$. These transitions
lead to unwanted mixing of the qubit states \cite{LZ-trans-SIF}.
In the vicinity of point $A'$, where the gap between the first and
second (not shown in Fig.\,2) energy bands is smaller
\cite{Abr-Steg}, $\hbar\Omega_{12} = (j_1-j_2)^2E_{J0}^2/2E_c$,
the Landau-Zener transitions $|1\rangle \leftrightarrow
|2\rangle$ are more probable. Fortunately, the second energy band
has the positive curvature,
$\frac{\partial^2E_2(0,\phi=\pi)}{\partial \phi^2}>0$, so the
mixing of these states might even improve the distinguishability
of signals from the ground and excited states. More rigorous
analysis of this effect on operation of the qubit in point $A$
is, however, needed.

Finally, let us evaluate the time of measurement required for the
resolution of the states $n=0$ and $n=1$ in the most favorable
magic points $A$ and $B$. As schematically shown in Fig.\,6, an rf
drive pulse is applied to the tank circuit just after manipulation
of the qubit ($t=0$) and induces growing oscillations in the tank.
The amplitude of the oscillations of voltage $V$ approaches a
steady value $A_0$ $(A_1)$ for $n=0$ $(n=1)$ after the time
$t_{\rm rise} \approx 2\pi Q/\omega_0$. Assuming a corresponding
amplitude of oscillations of phase $\phi_a=\pi/2$, we obtain for
the parameters of Table~1 the amplitudes
\begin{equation}
\label{V-out} A_0=\phi_a\frac{\Phi_0}{2\pi}\frac{\omega
L_T}{M}=\left(\frac{\pi \Phi_0 \omega R_p I_{c0}}{8k^2 Q
\beta_L}\right)^{1/2}\approx 3\;\mu{\rm V}
\end{equation}
and $A_1 \approx 1\:\mu$V.

Assuming that the equivalent noise of a semiconductor-based
amplifier referred to the input is of the order of Johnson voltage
noise across $R_p \approx 10$\,k$\Omega$ at ambient temperature
$T^*\sim 2$\,K, i.e., $S_V^{1/2} \approx 1$\,nV/$\sqrt{\rm Hz}$,
we can express the signal-to-noise ratio as
\begin{equation}
\label{SNR} \textrm{SNR}=\frac{(A_0-A_1)\sqrt{t_{\rm
meas}}}{S_V^{1/2}}\approx 2\:10^3 \sqrt{t_{\rm
meas}/1\,\textrm{s}},
\end{equation}
where $t_{\rm meas}$ is the time of measurement. This time should
clearly be much shorter than the relaxation time $\tau_{r}^{(p)}$
(evaluated as $\approx 0.1\,$s, see Table 1) and exceed the rise
time of the oscillations in the tank $t_{\rm rise}\approx
1\,\mu$s (the latter condition nicely agrees with the requirement
SNR $>1$). Thus, a drive pulse duration of $\sim 10\,\mu$s
ensuring $t_{\rm meas} \sim 10\,\mu$s seems to be a good choice
as it yields the sufficiently high value of SNR\,$\approx 6$. The
latter (as well as the quantum quality factor $Q_{\varphi}$) can
be substantially improved by using a SQUID-based low-noise
amplifier \cite{Mueck}.

\section{Conclusion}

We have demonstrated that both manipulation and readout of the
charge-phase qubit coupled to a tank circuit is, in principle,
possible. More specifically, the decoherence effect of the
electric and magnetic control lines can seemingly be minimized by
reducing coupling to the qubit. The readout system based on the
parallel tank circuit and cold amplifier can ensure sufficiently
weak dephasing in the regime without rf drive. The dephasing rate
strongly depends on the accuracy of adjusting the offset flux bias
$\Phi_m = 0$ or $\Phi_m =\Phi_0/2$ corresponding to operation in
the magic points. High symmetry of the Josephson junction
parameters may further improve the coherence characteristics of
the qubit. Since the $LC$ resonance tank circuit introduces only
small noise at the high transition frequency of the qubit,
$\Omega\gg \omega_0$, the rate of relaxation can also be made
sufficiently small. Applying an rf drive pulse of limited span
allows a readout of the qubit state in the regimes of single and
repeated measurements.

Other problems in engineering Josephson qubits with weak
decoherence are the $1/f$ noise of critical currents of Josephson
junctions \cite{vanHarlingen} and the $1/f$ background noise
coupled to the charge variable \cite{Paladino} which have not been
addressed in this paper but are equally important. Hopefully, in
the given system these effects might not be as strong as in
"traditional" tunnel-junction devices like dc SQUIDs and single
electron transistors operating at nonzero voltage bias. Due to
perfect decoupling of the superconducting loop with the single
charge transistor from dc bias lines and due to the coherent
nature of the Josephson current in the zero voltage bias regime,
one could expect a minor backaction effect of the zero-bias
operating transistor on its critical current noise and charge
noise which dramatically depend on the current fed (see, for
example, Ref.\,\cite{Krupenin}).

\section{Acknowledgements}

The author would like to thank Per Delsing, Yuriy Makhlin and
Frank Wilhelm for stimulating discussions. This work was
partially supported by the European Union through the SQUBIT-2
project.



\begin{references}
\bibitem{Quantr} In fact, the energy spectrum of this system is similar
to that of Quantronium \cite{Vion}, having one additional large
Josephson junction in the loop.
\end{references}
\end{document}